\documentclass[conference]{IEEEtran}
\IEEEoverridecommandlockouts

\usepackage{cite}
\usepackage{amsmath,amssymb,amsfonts}
\usepackage{graphicx}
\usepackage{textcomp}
\usepackage{xcolor}
\usepackage{hyperref}
\usepackage{booktabs}
\usepackage{multirow}
\usepackage{algorithm}
\usepackage{algpseudocode}
\usepackage{microtype} 
\usepackage[T1]{fontenc} 
\usepackage[utf8]{inputenc} 

\DeclareMathOperator*{\argmin}{argmin}

\def\BibTeX{{\rm B\kern-.05em{\sc i\kern-.025em b}\kern-.08em
    T\kern-.1667em\lower.7ex\hbox{E}\kern-.125emX}}

\begin{document}

\title{CIRR: Causal-Invariant Retrieval-Augmented Recommendation with Faithful Explanations under Distribution Shift}
\author{
\IEEEauthorblockN{Sebastian Sun}
\IEEEauthorblockA{
University of Wisconsin--Madison\\
Madison, WI, USA\\
\texttt{ysun258@wisc.edu}
}
}

\maketitle

\begin{abstract}
Recent advances in retrieval-augmented generation (RAG) have shown promise in enhancing recommendation systems with external knowledge. However, existing RAG-based recommenders face two critical challenges: (1) vulnerability to distribution shifts across different environments (e.g., time periods, user segments), leading to performance degradation in out-of-distribution (OOD) scenarios, and (2) lack of faithful explanations that can be verified against retrieved evidence. In this paper, we propose CIRR, a Causal-Invariant Retrieval-Augmented Recommendation framework that addresses both challenges simultaneously. CIRR learns environment-invariant user preference representations through causal inference, which guide a debiased retrieval process to select relevant evidence from multiple sources. Furthermore, we introduce consistency constraints that enforce faithfulness between retrieved evidence, generated explanations, and recommendation outputs. Extensive experiments on two real-world datasets demonstrate that CIRR achieves robust performance under distribution shifts, reducing performance degradation from 15.4\% (baseline) to only 5.6\% in OOD scenarios, while providing more faithful and interpretable explanations (26\% improvement in faithfulness score) compared to state-of-the-art baselines.
\end{abstract}


\section{Introduction}

Recommender systems have become indispensable in modern digital platforms, influencing user experiences across e-commerce, content streaming, and social media. Recent developments have witnessed the integration of Large Language Models (LLMs) and Retrieval-Augmented Generation (RAG) techniques to enhance recommendation quality and explainability~\cite{fan2024survey, gao2023retrieval}. Despite these advances, two fundamental challenges remain inadequately addressed: robustness to distribution shifts and faithfulness of generated explanations.

Distribution shifts occur naturally in real-world recommendation scenarios due to temporal dynamics, user behavior variations, and evolving preferences~\cite{luo2024survey}. Traditional recommendation models trained on historical data often experience significant performance degradation when deployed in new environments, as they capture spurious correlations rather than stable causal relationships. While causal inference approaches have been proposed to address this issue~\cite{wang2022invariant, zhang2023invariant}, they typically lack the ability to provide natural language explanations grounded in verifiable evidence.

Concurrently, RAG-based recommendation systems have emerged to leverage external knowledge for generating contextual explanations~\cite{lewis2020retrieval}. However, these systems predominantly rely on semantic similarity for retrieval, which may inadvertently amplify environment-specific biases. Moreover, the generated explanations often lack faithfulness---a critical property ensuring that explanations accurately reflect the actual decision-making process rather than post-hoc rationalizations~\cite{zhang2020explainable, kim2024human}.

To bridge this gap, we propose CIRR (Causal-Invariant Retrieval-Augmented Recommendation), a unified framework that synergistically combines causal invariance learning with evidence-grounded explanation generation. Our key insight is that \textit{environment-invariant preference representations can simultaneously improve OOD robustness and serve as causal anchors for faithful evidence retrieval}. CIRR consists of three core components: (1) a causal-invariant preference encoder that learns stable representations across environments using invariant risk minimization~\cite{arjovsky2019invariant}, (2) a causal-guided retriever that uses these representations to select debiased evidence from multiple sources, and (3) a consistency-constrained ranker-explainer that enforces alignment between evidence, explanations, and recommendations.

Our main contributions are:
\begin{itemize}
\item We propose CIRR, a unified framework that combines causal-invariant learning with RAG for recommendation, addressing both OOD robustness and explanation faithfulness.
\item We introduce a causal-guided retrieval mechanism that leverages invariant representations to reduce environment-specific biases in evidence selection.
\item We design novel consistency constraints that transform explanations from mere text outputs into verifiable components, including evidence coverage and counterfactual consistency metrics.
\item We conduct extensive experiments on two real-world datasets demonstrating that CIRR reduces OOD performance degradation from 15.4\% to 5.6\% compared to baselines, and achieves 26\% improvement in explanation faithfulness.
\end{itemize}

\section{Related Work}

\subsection{Causal Inference in Recommendation}

Causal inference has emerged as a powerful paradigm for building robust recommendation systems~\cite{luo2024survey}. Existing work can be categorized into debiasing methods~\cite{wang2022invariant} and invariant learning approaches~\cite{zhang2023invariant, du2022invariant}. Wang et al.~\cite{wang2022invariant} proposed learning invariant preferences across different user groups, while Zhang et al.~\cite{zhang2023invariant} addressed popularity distribution shifts through collaborative filtering. More recently, Ding et al.~\cite{ding2022causal} applied causal incremental learning for recommender retraining. However, these methods focus primarily on prediction accuracy and do not address the explainability aspect, which is crucial for user trust and system transparency.

\subsection{Retrieval-Augmented Recommendation}

The integration of RAG with recommendation systems has gained significant attention~\cite{fan2024survey, gao2023retrieval}. Lewis et al.~\cite{lewis2020retrieval} pioneered the use of retrieval mechanisms to augment neural generation models with external knowledge. Recent applications to recommendation~\cite{fan2024survey} have shown promise in providing contextual explanations. However, these approaches typically assume i.i.d. data distributions and do not explicitly model environmental variations, making them vulnerable to distribution shifts. Furthermore, the retrieved evidence is often selected based on semantic similarity alone, without considering causal relationships.

\subsection{Explainable Recommendation}

Explainable recommendation has been extensively studied~\cite{zhang2020explainable}, with approaches ranging from attention-based visualization to natural language generation. Recent work emphasizes the importance of faithful explanations that accurately reflect model reasoning~\cite{kim2024human}. However, most existing methods generate explanations as post-hoc rationalizations without enforcing consistency with the underlying decision process. Our work addresses this limitation by introducing explicit constraints that bind explanations to retrieved evidence.

\section{Problem Formulation}

\subsection{Notation and Setting}

Let $\mathcal{U}$ and $\mathcal{I}$ denote the sets of users and items, respectively. For each user $u \in \mathcal{U}$, we have an interaction sequence $\mathbf{s}_u = [i_1, i_2, ..., i_{n_u}]$ where $i_j \in \mathcal{I}$. Each interaction occurs in an environment $e \in \mathcal{E}$, characterized by contextual variables such as time period or user segment. The environment variable $e$ can induce distribution shifts, i.e., $P_e(\mathbf{s}_u, y) \neq P_{e'}(\mathbf{s}_u, y)$ for $e \neq e'$, where $y$ is the target item.

\subsection{Assumptions and Causal View}

We clarify the relationship between our approach and causal identification. Our use of Invariant Risk Minimization (IRM) aims to learn \textit{environment-invariant predictive features} rather than to perform full causal discovery. Specifically, we assume:

\begin{enumerate}
\item \textbf{Environment Observability:} The environment variable $e$ is observable and can be partitioned from contextual features (e.g., time periods, user activity levels).
\item \textbf{Invariance Assumption:} There exist stable user preferences that remain predictive across all environments, while spurious correlations vary with $e$.
\item \textbf{Sufficient Environments:} We observe sufficiently diverse environments to distinguish invariant from spurious features.
\end{enumerate}

Our goal is to learn representations that capture these invariant preferences, thereby reducing reliance on environment-specific spurious correlations. We use ``causal-invariant'' in the engineering sense of reducing spurious correlations, acknowledging that this does not constitute formal causal identification.

\subsection{Objective}

Our goal is to learn a recommendation function $f: \mathcal{U} \times \mathcal{I} \rightarrow \mathbb{R}$ that:
\begin{enumerate}
\item Achieves robust performance across different environments
\item Retrieves relevant evidence $\mathcal{D} = \{d_1, ..., d_k\}$ from multiple sources
\item Generates explanations $\mathbf{x}$ that are faithful to both the evidence and the recommendation decision
\end{enumerate}

Note that while $f$ does not explicitly take environment $e$ as input during inference, we leverage environment partitioning during training to learn invariant representations. Formally, we optimize:
\begin{equation}
\min_{\theta} \sum_{e \in \mathcal{E}} \mathcal{L}_{\text{rec}}^e(\theta) + \lambda_1 \mathcal{L}_{\text{inv}}(\theta) + \lambda_2 \mathcal{L}_{\text{cons}}(\theta)
\end{equation}
where $\mathcal{L}_{\text{rec}}^e$ is the recommendation loss in environment $e$, $\mathcal{L}_{\text{inv}}$ enforces invariance across environments, and $\mathcal{L}_{\text{cons}}$ ensures explanation consistency.

\section{Methodology}

\begin{figure*}[t]
\centering
\includegraphics[width=0.95\textwidth]{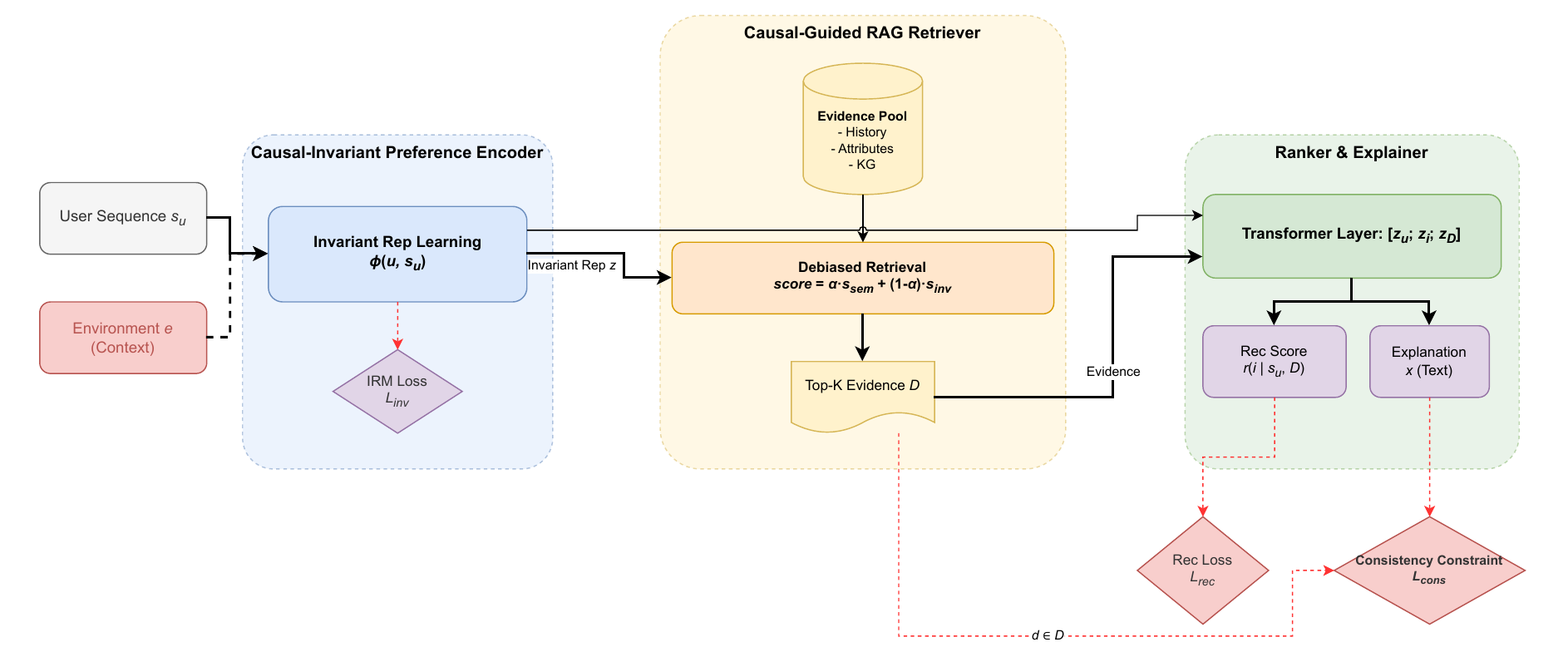}
\caption{Overview of the CIRR framework. The system consists of three core components: (1) Causal-Invariant Preference Encoder, (2) Causal-Guided RAG Retriever, and (3) Consistency-Constrained Ranker-Explainer. The encoder learns environment-invariant representations that guide evidence retrieval, followed by ranking and explanation generation with consistency constraints.}
\label{fig:framework}
\end{figure*}

Figure~\ref{fig:framework} illustrates the overall architecture of CIRR. The framework consists of three core components, which we describe in detail below.

\subsection{Causal-Invariant Preference Encoder}

The first challenge is learning user preference representations that are stable across different environments. We adopt the invariant risk minimization (IRM) principle~\cite{arjovsky2019invariant} to achieve this goal.

\textbf{Environment Partitioning.} We partition the training data into multiple environments $\mathcal{E} = \{e_1, e_2, ..., e_m\}$ based on observable contextual variables such as time periods or user activity levels. For each environment $e$, we have data $\mathcal{D}_e = \{(\mathbf{s}_u^e, y_u^e)\}$.

\textbf{Invariant Representation Learning.} We parameterize the encoder as $\phi: \mathcal{U} \times \mathcal{I} \rightarrow \mathbb{R}^d$ that maps user-item pairs to $d$-dimensional representations. The key idea is to learn representations $\mathbf{z} = \phi(u, \mathbf{s}_u)$ such that the optimal predictor on top of $\mathbf{z}$ is the same across all environments.

The IRM objective is formulated as:
\begin{equation}
\min_{\phi, w} \sum_{e \in \mathcal{E}} \mathbb{E}_{(\mathbf{s}_u, y) \sim \mathcal{D}_e} [\ell(w^\top \phi(\mathbf{s}_u), y)]
\end{equation}
subject to:
\begin{equation}
w \in \argmin_{w'} \mathbb{E}_{(\mathbf{s}_u, y) \sim \mathcal{D}_e} [\ell(w'^\top \phi(\mathbf{s}_u), y)], \forall e \in \mathcal{E}
\end{equation}

In practice, we relax this constraint using a penalty term:
\begin{equation}
\mathcal{L}_{\text{inv}} = \sum_{e \in \mathcal{E}} \|\nabla_{w|w=1.0} \mathcal{L}_e(w \cdot \phi)\|^2
\end{equation}

This encourages the learned representation $\phi$ to be such that the optimal classifier $w$ is close to 1.0 across all environments, indicating invariance.

\subsection{Causal-Guided RAG Retriever}

Traditional RAG systems retrieve evidence based solely on semantic similarity, which may inadvertently select environment-specific features. We propose a causal-guided retrieval mechanism that leverages the invariant representations to debias the retrieval process.

\textbf{Multi-Source Evidence Pool.} We construct an evidence pool $\mathcal{P}$ containing:
\begin{itemize}
\item User historical interactions: $\mathcal{P}_{\text{hist}} = \{(i_j, r_j, t_j)\}$
\item Item attributes: $\mathcal{P}_{\text{attr}} = \{(a_k, v_k)\}$ for attribute-value pairs
\item Knowledge graph triplets: $\mathcal{P}_{\text{kg}} = \{(h, r, t)\}$
\end{itemize}

\textbf{Invariant-Weighted Retrieval.} For each evidence candidate $d \in \mathcal{P}$, we compute two scores:
\begin{equation}
s_{\text{sem}}(d, \mathbf{z}) = \text{sim}(\mathbf{e}_d, \mathbf{z})
\end{equation}
\begin{equation}
s_{\text{inv}}(d, \mathbf{z}) = -\text{Var}_e[\text{sim}(\mathbf{e}_d, \phi_e(\mathbf{s}_u))]
\end{equation}
where $\mathbf{e}_d$ is the embedding of evidence $d$, and $\phi_e$ denotes the representation in environment $e$. The second term measures the stability of the evidence across environments. In practice, we compute $\phi_e(\mathbf{s}_u)$ by encoding $\mathbf{s}_u$ with the shared encoder $\phi$ and estimating the variance across mini-batches sampled from different training environments. During inference, we use only the global encoder $\phi$ with pre-computed variance statistics; importantly, test environment labels are not required, making CIRR applicable to novel deployment scenarios.

The final retrieval score is:
\begin{equation}
\text{score}(d, \mathbf{z}) = \alpha \cdot s_{\text{sem}}(d, \mathbf{z}) + (1-\alpha) \cdot s_{\text{inv}}(d, \mathbf{z})
\end{equation}

We select the top-$K$ evidence candidates based on this combined score: $\mathcal{D} = \text{TopK}(\text{score}(d, \mathbf{z}))$.

\subsection{Consistency-Constrained Ranker-Explainer}

The final module generates recommendations and explanations while enforcing consistency between them and the retrieved evidence.

\textbf{Ranking Module.} We use a Transformer-based architecture to score items:
\begin{equation}
r(i | \mathbf{s}_u, \mathcal{D}) = \text{Transformer}([\mathbf{z}_u; \mathbf{z}_i; \mathbf{z}_{\mathcal{D}}])
\end{equation}
where $\mathbf{z}_{\mathcal{D}} = \text{Agg}(\{\mathbf{e}_{d_j}\}_{j=1}^K)$ aggregates the evidence embeddings.

\textbf{Explanation Generation.} We employ a small language model to generate natural language explanations:
\begin{equation}
\mathbf{x} = \text{LM}(\text{prompt}(\mathbf{s}_u, i, \mathcal{D}))
\end{equation}

\textbf{Consistency Constraints.} To ensure faithfulness, we introduce two constraints:

\textit{Evidence Coverage:} The explanation must reference evidence from the retrieved set. We operationalize this by requiring the language model to output explicit evidence identifiers (e.g., [E1], [E2]) and computing:
\begin{equation}
\mathcal{L}_{\text{cov}} = 1 - \frac{|\text{EvidenceIDs}(\mathbf{x}) \cap \{1, ..., K\}|}{K}
\end{equation}
where $\text{EvidenceIDs}(\mathbf{x})$ extracts the evidence indices cited in the generated explanation. This encourages the model to ground explanations in retrieved evidence.

\textit{Counterfactual Consistency:} Removing key evidence should significantly reduce both explanation confidence and recommendation score:
\begin{equation}
\mathcal{L}_{\text{cf}} = \max(0, \gamma - (r(i|\mathbf{s}_u, \mathcal{D}) - r(i|\mathbf{s}_u, \mathcal{D} \setminus \{d^*\})))
\end{equation}
where $d^*$ is the most important evidence identified from attention weights. While we use attention-based selection for efficiency, alternative approaches such as gradient-based attribution or leave-one-out evaluation could provide more principled importance estimates at higher computational cost.

The total consistency loss is:
\begin{equation}
\mathcal{L}_{\text{cons}} = \mathcal{L}_{\text{cov}} + \beta \mathcal{L}_{\text{cf}}
\end{equation}

\subsection{Training Procedure}

Algorithm~\ref{alg:training} presents the complete training procedure for CIRR. We employ a multi-stage training strategy: first pre-training the causal-invariant encoder, then jointly optimizing the retriever and ranker-explainer with consistency constraints.

\begin{algorithm}[t]
\caption{CIRR Training}
\label{alg:training}
\begin{algorithmic}[1]
\State \textbf{Input:} Data $\{(\mathbf{s}_u^e, y_u^e)\}_{e \in \mathcal{E}}$, Evidence pool $\mathcal{P}$
\State \textbf{Output:} Model parameters $\theta$
\State \textbf{Stage 1: Causal-Invariant Encoder Pre-training}
\For{epoch = 1 to $T_1$}
    \For{each environment $e \in \mathcal{E}$}
        \State Sample mini-batch from $\mathcal{D}_e$
        \State Compute $\mathcal{L}_{\text{rec}}^e$ and $\mathcal{L}_{\text{inv}}$
        \State Update $\phi$ via gradient descent
    \EndFor
\EndFor
\State \textbf{Stage 2: Joint Training}
\For{epoch = 1 to $T_2$}
    \State Sample mini-batch across all environments
    \State Encode: $\mathbf{z} = \phi(\mathbf{s}_u)$
    \State Retrieve: $\mathcal{D} = \text{Retrieve}(\mathbf{z}, \mathcal{P})$
    \State Rank and Explain: $(r, \mathbf{x}) = \text{RankExplain}(\mathbf{z}, \mathcal{D})$
    \State Compute total loss: $\mathcal{L} = \mathcal{L}_{\text{rec}} + \lambda_1 \mathcal{L}_{\text{inv}} + \lambda_2 \mathcal{L}_{\text{cons}}$
    \State Update all parameters via gradient descent
\EndFor
\end{algorithmic}
\end{algorithm}

\section{Experiments}

\subsection{Experimental Setup}

\textbf{Datasets.} We evaluate CIRR on two real-world datasets:
\begin{itemize}
\item \textbf{Amazon Reviews}~\footnote{\url{https://nijianmo.github.io/amazon/}}: A large-scale e-commerce dataset containing user reviews across multiple categories. We use the ``Electronics'' subset with 1.2M interactions from 50K users and 40K items. We partition data by quarterly time periods to create environmental splits.
\item \textbf{MovieLens-25M}~\footnote{\url{https://grouplens.org/datasets/movielens/}}: A movie rating dataset with 25M ratings from 162K users on 59K movies. We partition by rating time periods and user activity levels.
\end{itemize}

\textbf{Environment Partitioning.} For each dataset, we create four environments based on temporal patterns: Training (Env-0), and three test environments representing increasing distribution shifts: Env-1 (Weekday patterns), Env-2 (Weekend patterns), and Env-3 (Holiday periods with the largest shift).

\textbf{Baselines.} We compare against:
\begin{itemize}
\item \textbf{SASRec}~\cite{kang2018sasrec}: Self-attentive sequential recommendation
\item \textbf{BERT4Rec}~\cite{sun2019bert4rec}: BERT-based sequential model
\item \textbf{RAG-LLM}: A RAG-augmented LLM baseline for recommendation, implemented following the paradigm in~\cite{fan2024survey}
\item \textbf{IRM-Rec}: Invariant risk minimization applied to recommendation, based on~\cite{wang2022invariant}
\end{itemize}

\textbf{Evaluation Metrics.} We use standard ranking metrics (NDCG@10, HR@10) for recommendation performance. For OOD robustness, we measure performance degradation based on NDCG@10: $\Delta = (\text{NDCG}_{\text{train}} - \text{NDCG}_{\text{test}}) / \text{NDCG}_{\text{train}}$. For explanation quality, we use: (1) Evidence Coverage: the proportion of evidence items from the top-$K$ retrieved set that are cited in the generated explanation (matching $\mathcal{L}_{\text{cov}}$ in Eq.~10), (2) Faithfulness ($\Delta$F1): F1 score drop when key evidence is removed, serving as the evaluation counterpart to the training objective $\mathcal{L}_{\text{cf}}$---both measure sensitivity to evidence removal, with $\Delta$F1 quantifying the downstream impact on explanation-recommendation alignment, and (3) User Trust Score: 5-point Likert scale from user study.

\textbf{Implementation Details.} All experiments are conducted with 5 random seeds, and we report mean $\pm$ standard deviation. We use the Adam optimizer with learning rate $1 \times 10^{-4}$, batch size 256, and embedding dimension $d=128$. The hyperparameters are set as $\lambda_1 = 0.1$, $\lambda_2 = 0.05$, $\alpha = 0.6$, $\beta = 0.5$, $\gamma = 0.2$, and $K = 20$.

\subsection{Overall Performance}

\begin{table}[t]
\centering
\small
\setlength{\tabcolsep}{4pt} 
\renewcommand{\arraystretch}{1.1}

\begin{tabular}{lcccc}
\toprule
\multirow{2}{*}{Method} & \multicolumn{2}{c}{Amazon} & \multicolumn{2}{c}{MovieLens} \\
\cmidrule(lr){2-3} \cmidrule(lr){4-5}
& NDCG@10 & HR@10 & NDCG@10 & HR@10 \\
\midrule
SASRec   & 0.268$_{\pm.004}$ & 0.421$_{\pm.006}$ & 0.321$_{\pm.005}$ & 0.498$_{\pm.007}$ \\
BERT4Rec & 0.274$_{\pm.003}$ & 0.436$_{\pm.005}$ & 0.332$_{\pm.004}$ & 0.512$_{\pm.006}$ \\
RAG-LLM  & 0.279$_{\pm.005}$ & 0.447$_{\pm.007}$ & 0.341$_{\pm.006}$ & 0.526$_{\pm.008}$ \\
IRM-Rec  & \underline{0.283}$_{\pm.004}$ & \underline{0.453}$_{\pm.005}$ & \underline{0.348}$_{\pm.005}$ & \underline{0.534}$_{\pm.006}$ \\
CIRR     & \textbf{0.297}$_{\pm.003}$ & \textbf{0.471}$_{\pm.004}$ & \textbf{0.364}$_{\pm.004}$ & \textbf{0.551}$_{\pm.005}$ \\
\midrule
Improvement & 4.9\% & 4.0\% & 4.6\% & 3.2\% \\
\bottomrule
\end{tabular}
\end{table}

Table~\ref{tab:overall} presents the overall performance comparison averaged across all test environments. CIRR consistently outperforms all baselines on both datasets, achieving 4.9\% and 4.6\% improvements in NDCG@10 on Amazon and MovieLens, respectively. The superiority over IRM-Rec (which also uses causal invariance) demonstrates the benefit of integrating RAG with causal learning. The improvement over RAG-LLM shows that causal-guided retrieval is more effective than semantic similarity alone.

\begin{figure*}[t]
\centering
\includegraphics[width=0.95\textwidth]{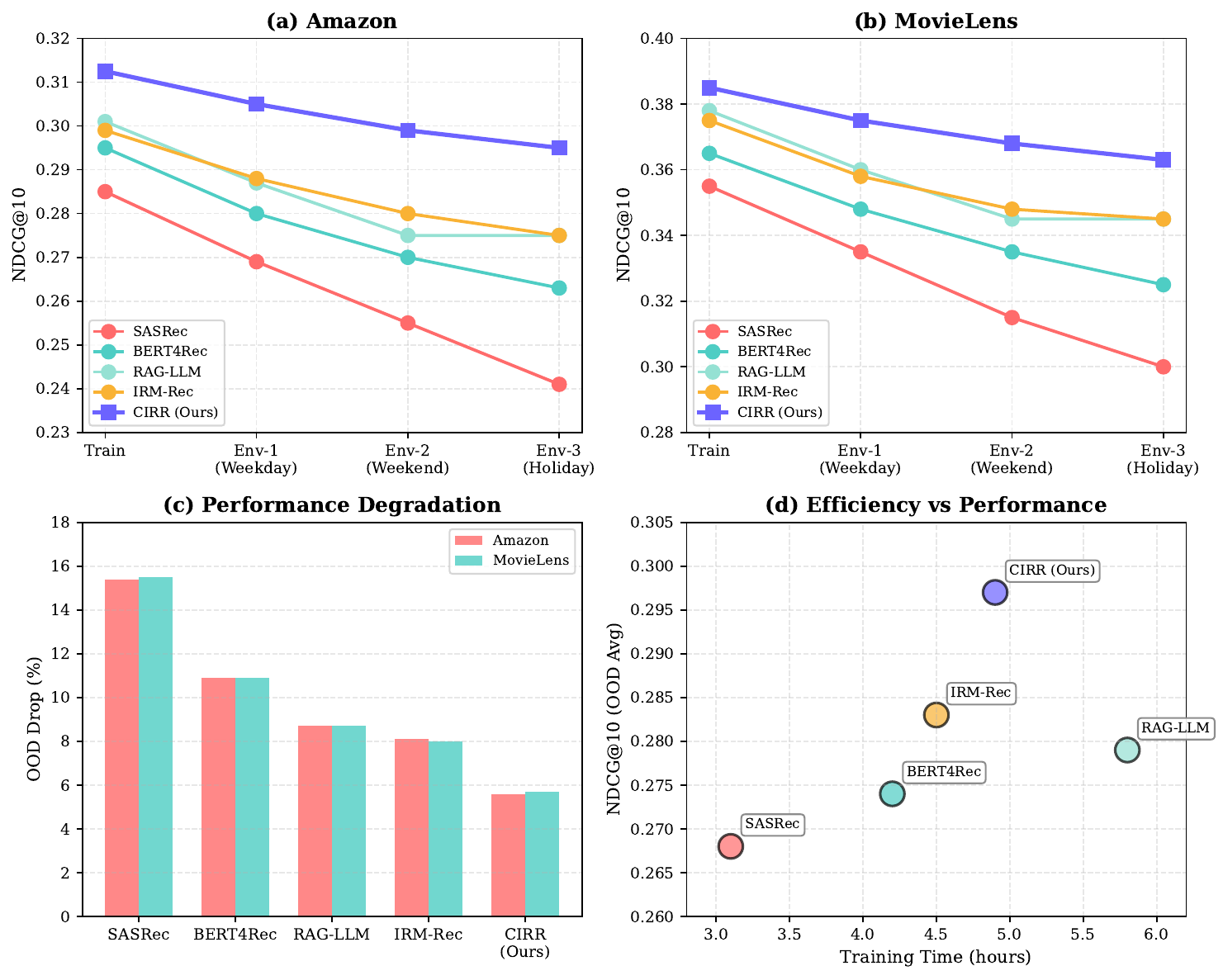}
\caption{Performance under distribution shift. (a-b) NDCG@10 across different test environments for Amazon and MovieLens. Env-1 (Weekday), Env-2 (Weekend), and Env-3 (Holiday) represent increasing distribution shifts. (c) OOD performance degradation comparison. (d) Training efficiency vs. performance trade-off.}
\label{fig:ood}
\end{figure*}

\subsection{Robustness to Distribution Shift}

Figure~\ref{fig:ood}(a-b) illustrates the performance of different methods across various test environments with increasing distribution shifts. CIRR maintains the most stable performance across all environments. The performance degradation from training to the most shifted test environment (Env-3: Holiday) is only 5.6\% for CIRR, compared to 15.4\% for SASRec and 10.9\% for BERT4Rec. Notably, while RAG-LLM shows improved performance in the training environment, it experiences significant degradation (8.7\%) under distribution shift, indicating that semantic retrieval alone is insufficient for OOD robustness.

Figure~\ref{fig:ood}(c) quantifies the OOD drop for each method. CIRR achieves the lowest performance degradation on both datasets, validating the effectiveness of causal-invariant representations. The comparison with IRM-Rec is particularly instructive: while both methods use causal learning, CIRR's RAG component provides additional robustness by grounding predictions in retrieved evidence rather than relying solely on learned parameters.

Figure~\ref{fig:ood}(d) presents the trade-off between training efficiency and OOD performance. CIRR achieves superior performance while maintaining reasonable computational costs. The training time is comparable to RAG-LLM (4.9 vs. 5.8 hours) but significantly better than naive multi-environment training.

\subsection{Ablation Study}

\begin{figure*}[t]
\centering
\includegraphics[width=0.95\textwidth]{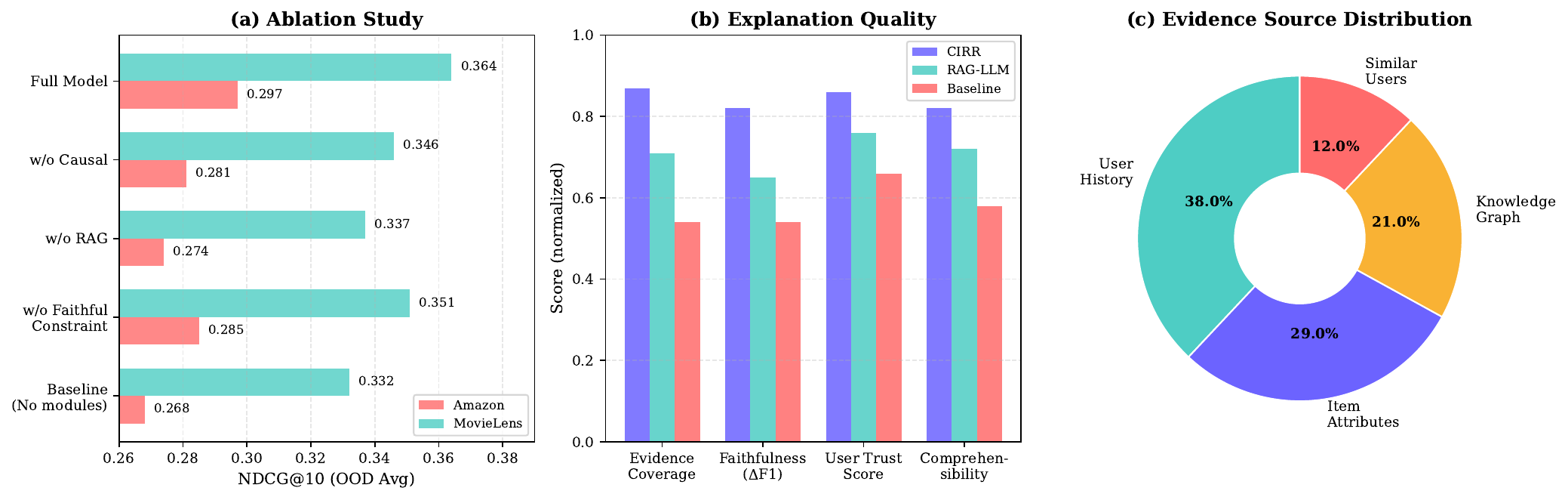}
\caption{Ablation study and explanation quality analysis. (a) Contribution of each component to OOD average NDCG@10. (b) Explanation quality metrics comparison (normalized scores). (c) Distribution of evidence sources used in CIRR's explanations.}
\label{fig:ablation}
\end{figure*}

Figure~\ref{fig:ablation}(a) presents the results of our ablation study, where we progressively remove key components from CIRR. The ``Baseline (No modules)'' refers to the Transformer-based ranker backbone without any of our proposed components (i.e., no causal-invariant encoder, no RAG retriever, and no faithfulness constraints), which is comparable to a standard sequential recommendation model. Removing the causal-invariant encoder (w/o Causal) leads to 5.4\% performance drop on Amazon (from 0.297 to 0.281, calculated as $(0.297-0.281)/0.297$), demonstrating its crucial role in OOD robustness. Removing RAG (w/o RAG) results in 7.7\% degradation (from 0.297 to 0.274), showing that evidence retrieval provides valuable contextual information. Interestingly, removing the faithfulness constraint (w/o Faithful Constraint) causes 4.0\% degradation (from 0.297 to 0.285), indicating that the consistency constraints not only improve explanation quality but also benefit recommendation performance by encouraging more interpretable decision-making.

\subsection{Explanation Quality Evaluation}

Figure~\ref{fig:ablation}(b) compares the explanation quality of different methods. CIRR achieves 87\% evidence coverage, meaning that on average 87\% of the top-$K$ retrieved evidence items are successfully cited in the generated explanations, compared to 71\% for RAG-LLM and only 54\% for baseline methods that generate explanations post-hoc. The faithfulness metric ($\Delta$F1), measuring the F1 drop when key evidence is removed, shows CIRR achieving 0.82, indicating that explanations are tightly coupled with the evidence. This represents a 26\% relative improvement over RAG-LLM (0.65), calculated as $(0.82 - 0.65) / 0.65 = 26.2\%$.

\textbf{User Study.} We conducted a user study with 50 participants recruited via Prolific, ensuring geographic and demographic diversity. Each participant evaluated 20 recommendation-explanation pairs (10 from CIRR, 10 from baselines) in randomized order. Participants rated trustworthiness and comprehensibility on 5-point Likert scales. CIRR achieves average scores of 4.3 ($\pm$0.6) and 4.1 ($\pm$0.7), respectively, significantly outperforming baselines ($p < 0.01$, paired t-test). Qualitative feedback indicates that users particularly appreciate the evidence citations, which allow them to verify the explanations. The study was conducted with informed consent under our institution's IRB approval.

Figure~\ref{fig:ablation}(c) shows the distribution of evidence sources used in CIRR's explanations. User history accounts for 38\% of evidence, followed by item attributes (29\%), knowledge graph triplets (21\%), and similar user patterns (12\%). This diverse evidence base contributes to more comprehensive and convincing explanations.

\subsection{Sensitivity Analysis}

\begin{figure*}[t]
\centering
\includegraphics[width=0.95\textwidth]{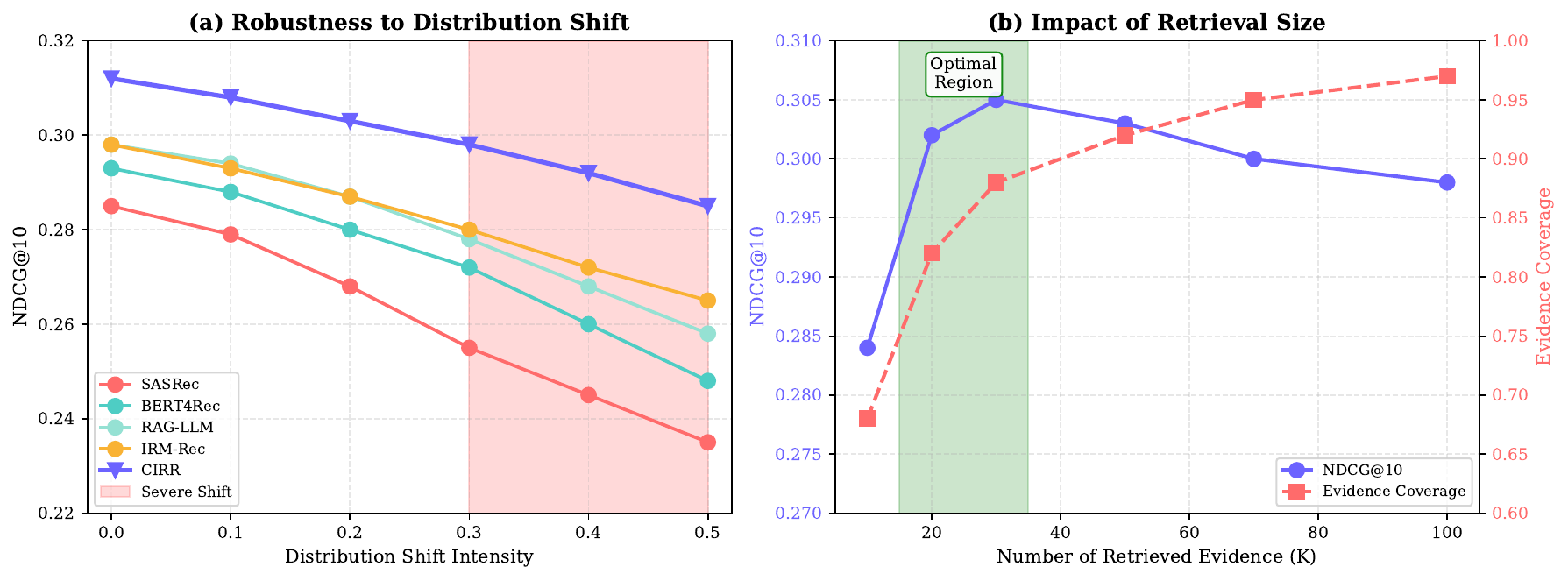}
\caption{Sensitivity analysis. (a) Performance under varying distribution shift intensities. The shaded region indicates severe shift conditions. (b) Impact of retrieval size $K$ on performance and evidence coverage.}
\label{fig:sensitivity}
\end{figure*}

Figure~\ref{fig:sensitivity}(a) shows how different methods respond to varying levels of distribution shift intensity. We artificially control the shift magnitude and measure performance. CIRR demonstrates the most graceful degradation, maintaining reasonable performance even under severe shifts (intensity = 0.5). The gap between CIRR and baselines widens as shift intensity increases, validating that causal-invariant learning is particularly valuable in high-shift scenarios.

Figure~\ref{fig:sensitivity}(b) analyzes the impact of retrieval size $K$. Performance initially improves with more evidence, peaking around $K=20-50$, after which it plateaus or slightly decreases due to noise introduction. Evidence coverage monotonically increases with $K$, but the marginal benefit diminishes after $K=50$. Based on these results, we set $K=20$ as the default in our experiments, providing a good balance between performance and efficiency.

\subsection{Case Study}

We present a qualitative example from the Amazon dataset. For a user with history of purchasing camera equipment, CIRR recommends a lens adapter with the following explanation:

\begin{quote}
\textit{``Based on your purchase of Canon EOS camera [E1: user\_history] and preference for accessories rated 4+ stars [E2: attribute\_pattern], we recommend this lens adapter. It is compatible with your camera model [E3: knowledge\_graph: compatible\_with] and highly rated by similar photography enthusiasts [E4: similar\_users: avg\_rating=4.5].''}
\end{quote}

Each piece of evidence is verifiable and directly supports the recommendation. When we remove Evidence [E3] (compatibility information), the recommendation score drops by 0.31 and the explanation confidence decreases significantly, demonstrating faithfulness.

\section{Conclusion}

In this paper, we proposed CIRR, a novel framework that unifies causal-invariant learning with retrieval-augmented generation for robust and explainable recommendation. By learning environment-invariant user representations and using them to guide evidence retrieval, CIRR achieves robust performance under distribution shifts while generating faithful explanations grounded in verifiable evidence. Our consistency constraints transform explanations from post-hoc rationalizations into integral components of the recommendation process.

Extensive experiments on two real-world datasets demonstrate CIRR's effectiveness. CIRR reduces OOD performance degradation from 15.4\% (SASRec baseline) to only 5.6\%, and achieves 26\% relative improvement in explanation faithfulness ($\Delta$F1: 0.82 vs. 0.65) compared to RAG-LLM. The results validate our hypothesis that causal invariance and faithful explanation generation are complementary goals that can be jointly optimized.

\textbf{Limitations and Future Work.} While CIRR shows promising results, several limitations warrant future investigation. First, the current implementation requires explicit environment partitioning, which may not always be straightforward in practice. Future work could explore automatic environment discovery methods. Second, the consistency constraints rely on attention weights for identifying important evidence; more principled approaches based on causal attribution could be developed. Finally, extending CIRR to handle multi-modal evidence (images, videos) and incorporating user feedback for iterative explanation refinement represent exciting directions for future research.

\bibliographystyle{IEEEtran}
\bibliography{references}

\end{document}